\documentclass[aps,prl,floatfix,twocolumn,superscriptaddress]{revtex4}
\usepackage{bm}
\usepackage{epsf}
\usepackage{amssymb}
\usepackage{amsmath}
\usepackage{graphicx}
\usepackage{rotating}
\usepackage{epsfig}
\usepackage{psfrag}
\usepackage{amsmath}
\usepackage{hyperref}
\usepackage{subfigure}

\usepackage{color}

\begin{document}

\newcommand{\cm}[1]{\textcolor{red}{\bf #1}}

\def\br{{\bf r}}
\newcommand{\Tr}{\mbox{Tr}}
\renewcommand{\dag}{\dagger}

\newcommand{\PD}[2]{\frac{\partial{#1}}{\partial{#2}}}
\newcommand{\DD}[2]{\frac{d{#1}}{d{#2}}}
\newcommand{\BK}[1]{\left[#1\right]}
\newcommand{\bk}[1]{\left(#1\right)}
\newcommand{\bra}[1]{\left\langle{#1}\right|}
\newcommand{\ket}[1]{\left|{#1}\right\rangle}
\newcommand{\lr}[1]{\left\langle#1\right\rangle}
\newcommand{\mc}{\mathcal}
\newcommand{\mr}{\mathrm}
\newcommand{\mb}{\mathbf}
\newcommand{\be}{\begin{equation}}
\newcommand{\ee}{\end{equation}}
\newcommand{\ba}{\begin{eqnarray}}
\newcommand{\ea}{\end{eqnarray}}

\newcommand {\apgt} {\ {\raise-.5ex\hbox{$\buildrel>\over\sim$}}\ }

\title{Ferromagnetic exchange, spin-orbit coupling and spiral magnetism at the LaAlO$_3$/SrTiO$_3$ interface}
\author{Sumilan Banerjee}
\affiliation{Department of Physics, The Ohio State University, Columbus, Ohio, 43210}
\author{Onur Erten}
\affiliation{Department of Physics, The Ohio State University, Columbus, Ohio, 43210}
\author{Mohit Randeria}
\affiliation{Department of Physics, The Ohio State University, Columbus, Ohio, 43210}
\date{\today}

\maketitle
{\bf The electronic properties of the polar interface between insulating oxides is a subject of 
great current interest~\cite{NHwang2004,SMannhart2006,AnnuTriscone2011}. 
An exciting new development is the observation of robust
magnetism~\cite{NatMatBrinkman2007,NatPhyAshoori2011,NatPhyMoler2011,NatCommAriando2011,PRLChandrasekhar2011}
at the interface of two non-magnetic materials LaAlO$_3$ (LAO) and SrTiO$_3$ (STO). 
Here we present a microscopic theory for the formation
and interaction of  local moments, which depends on essential features of the LAO/STO interface.
We show that correlation-induced moments arise due to interfacial splitting of orbital
degeneracy.  We find that gate-tunable Rashba spin-orbit coupling at the interface 
influences the exchange interaction mediated by conduction electrons. 
We predict that the zero-field ground state is a long-wavelength spiral and show that its
evolution in an external field accounts semi-quantitatively for  
torque magnetometry data~\cite{NatPhyAshoori2011}. Our theory describes
qualitative aspects of the scanning SQUID measurements~\cite{NatPhyMoler2011} 
and makes several testable predictions for future experiments.
}

Recent experiments on the LAO/STO interface have seen tantalizing magnetic 
signals~\cite{NatMatBrinkman2007,NatPhyAshoori2011,NatPhyMoler2011,NatCommAriando2011,PRLChandrasekhar2011},
often persisting up to high temperatures $\sim 100$ K. 
A large magnetization of $0.3\!-\!0.4\mu_\mr{B}$ per interface Ti was
observed by torque measurements~\cite{NatPhyAshoori2011} in an external field.
In contrast, scanning SQUID experiments~\cite{NatPhyMoler2011} found an
inhomogenous state with a dense set of local moments, but no net magnetization
in most of the sample except for isolated micron-scale ferromagnetic patches.
Our goal is to reconcile these seemingly contradictory observations and 
to gain insight into the itinerant {\it versus} local moment nature of the magnetism, the exchange 
mechanism, and the ordered state. 
  
LAO and STO are both band insulators, but the TiO$_2$ layers at the 
interface are n-doped when LAO is terminated by a LaO$^{+}$ layer. 
The ``polar catastrophe''~\cite{NHwang2004} arising from a stack of charged LaO$^{+}$ and AlO$_2^{-}$ layers grown on STO
is averted by the the transfer of 0.5 electrons per interface Ti. In addition,
oxygen vacancies are also known to provide additional electrons at the interface~\cite{NHwang2004}. 

What is the fate of these electrons at the interface? Transport data suggests
that only a small fraction of the electrons (5-10\% of the 0.5 e/Ti) are mobile,
as estimated from the Hall effect and 
quantum oscillations~\cite{SMannhart2006,NatPhyMoler2011,PRLShalom2010, PRBDagan2011}.
Interestingly, magneto-transport studies show a large, and gate-tunable, 
Rashba spin-orbit coupling (SOC) for the conduction electrons,
arising from broken inversion at the interface~\cite{PRLCaviglia2010}. 
In fact, most of the electrons (comparable to 0.5 e/Ti) seem to behave like local moments 
in the magnetic measurements~\cite{NatPhyAshoori2011,NatPhyMoler2011} discussed 
above. 

In this paper we propose a microscopic model of electrons in the Ti  $t_{2g}$ states at 
the LAO/STO interface and show that it leads to the following results. 
(1) $S$=1/2 local moments form in the top TiO$_2$ layer due to Coulomb correlations, with interfacial
splitting of $t_{2g}$ degeneracy playing a critical role.
(2) Conduction electrons mediate ferromagnetic interactions between the 
moments via double-exchange. 
(3) Rashba SOC for the conduction electrons leads to a Dzyaloshinski-Moriya (DM) interaction
and a ``compass'' anisotropy term with a definite ratio of their strengths.
(4) The zero field ground state is a long wavelength spiral with a SOC-dependent pitch.
(5) The spiral transforms into a ferromagnetic state in an external field $H$.

We provide a semi-quantitative understanding of the torque magnetometry results for the magnetization $M(H)$.
We also reconcile some of the key differences between the torque and SQUID measurements and point to a novel magnetoelastic 
coupling effect that can be crucial for determining the magnetic ground state of this kind of polar interface. 
Our model naturally explains the coexistence of magnetism and superconductivity
at very low temperatures~\cite{NatPhyAshoori2011,NatPhyMoler2011,PRLChandrasekhar2011}.
Finally, we make a number of specific predictions that can be tested experimentally.

%--------------
{\bf Symmetry-based considerations:}
Many of our conclusions regarding magnetism can be understood qualitatively based on symmetry, 
provided we have local moments ${\bf S}_\mb{r}$ on a square lattice in the $x$-$y$ plane at the interface.
The detailed microscopic analysis given below shows how such a square lattice is formed and also gives 
quantitative insights into the parameter dependence of exchange couplings.
Symmetry dictates the form of the interactions, the first of which is an isotropic Heisenberg exchange 
$-J \sum_{{\bf r},\mu} {\bf S}_\mb{r}\cdot{\bf S}_{\mb{r}+b\hat{\mu}}$, where $\hat{\mu}=\hat{x},\hat{y}$
and $b$ the lattice spacing. The sign of $J$ is not determined by symmetry, but our microscopic
analysis leads to a ferromagnetic $J>0$. 
Inversion symmetry breaking at the interface implies that $\pm\hat{z}$ are not equivalent and
we can write down two SOC terms~\cite{PRMoriya1960}.
The DM term is {$D\sum_{{\bf r},\mu}\hat{d}_\mu$}$\cdot${$({\bf S}_\mb{r}$}$\times${${\bf S}_{\mb{r}+b\hat{\mu}})$} 
with {$\hat{d}_\mu =\hat{z}$}$\times${$\hat{\mu}$}, and 
the allowed compass anisotropy terms are $-\sum_{\bf r}[A'({S}_\mb{r}^x{S}_{\mb{r}+b\hat{x}}^x + {S}_\mb{r}^y{S}_{\mb{r}+b\hat{y}}^y)
+ A({S}_\mb{r}^x{S}_{\mb{r}+b\hat{y}}^x + {S}_\mb{r}^y{S}_{\mb{r}+b\hat{x}}^y)]$.

One can see, quite generally, that the ground state of such a model can be a long wavelength
spiral that looks locally ferromagnetic (FM), thus minimizing the $J$-term, but whose pitch is determined by the
small DM and compass terms. Our microscopic results for $D, A, A'$ unequivocally
predict such a spiral ground state for $H=0$. In an external magnetic field ($\gtrsim 1$T), however, we find a uniform FM state.
These observations permit us to understand the torque data in a field~\cite{NatPhyAshoori2011} semi-quantitatively,
and to see why over most of the sample the $H=0$ scanning SQUID measurements find no net moment~\cite{NatPhyMoler2011}.
We will discuss below detailed comparison with experiments and new predictions.

\begin{figure}[!t]
\includegraphics[width=8cm]{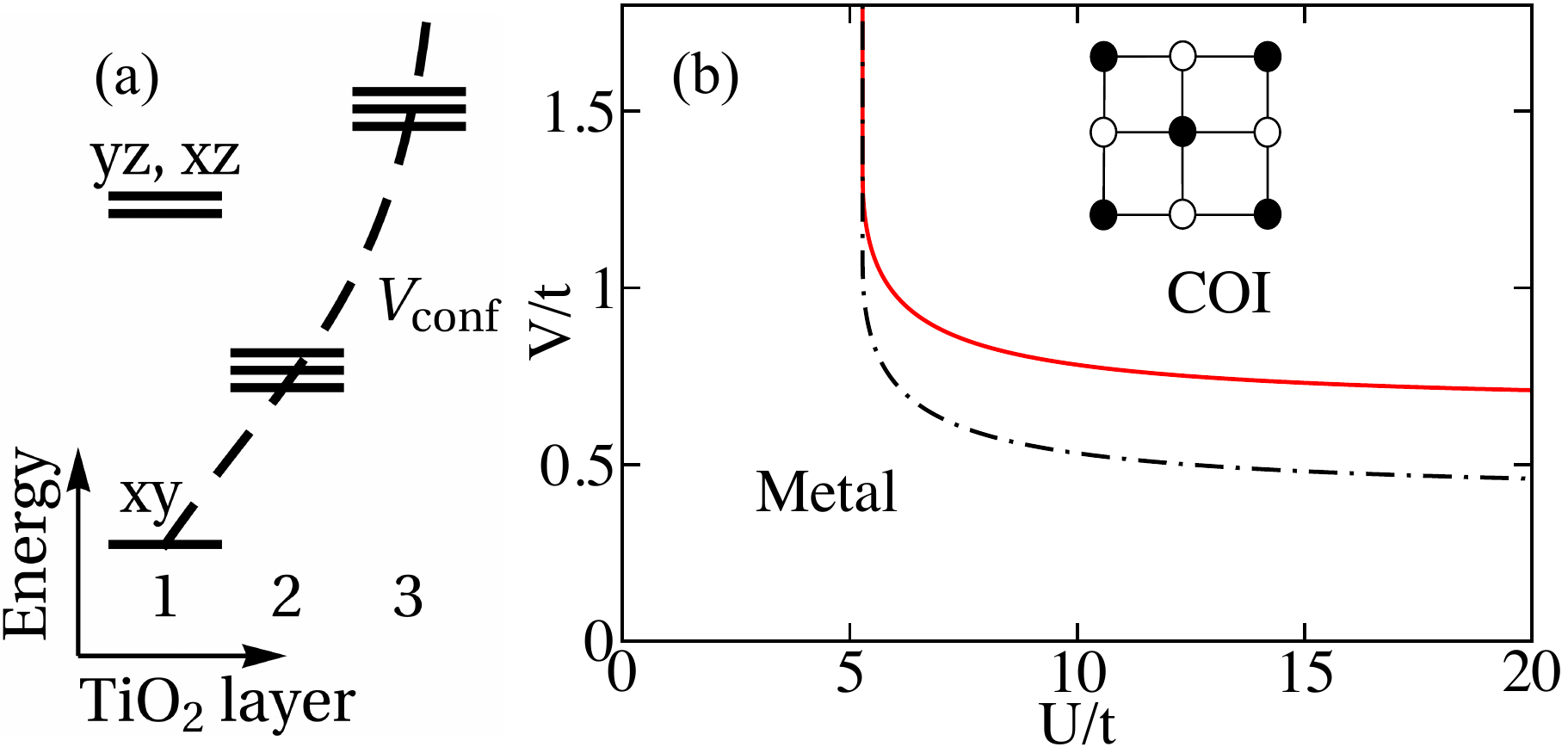}
\caption{{\bf Electronic orbitals and charge localization:} (a) Schematic of Ti $t_{2g}$ energy levels. 
At the interface (layer 1), the degeneracy is lifted with $d_{xy}$ lower than the $d_{yz}$ and $d_{xz}$ orbitals. 
The $t_{2g}$ energies in layers 2 and 3 increases due to the confining potential $V_{\rm conf}$. 
(b) Phase diagram of the single band extended Hubbard model at quarter-filing. 
A charge-ordered insulator (COI) is obtained for moderate values of 
correlations, on-site Hubbard $U$ and next-neighbor Coulomb $V$, above the solid red phase boundary. 
Coupling to the breathing mode phonon further stabilizes the COI. The dashed black line corresponds
to an energy gain $E_\mr{ph}=0.25t$ due to lattice distortion (see Methods).
\label{Fig:1}}
\end{figure}

{\bf Electronic structure:} Electrons at the interface are in Ti $t_{2g}$ states.
Both density functional theory (DFT)~\cite{PRBPickett2006}
and spectroscopic measurements \cite{PRLSalluzzo2009,PRLSing2009,PRBBerner2010},
show that, in the top TiO$_2$ layer [labeled ``1'' in Fig.~\ref{Fig:1}(a)], $d_{xy}$ states have lower energy 
than $d_{xz},d_{yz}$. In addition to $z$-confinement 
raising the $xz, yz$ levels, the mismatch of in-plane LAO and STO lattice parameters
leads to an out-of-plane distortion that lowers $d_{xy}$.  
Further, the $d_{xy}$ orbitals delocalize primarily in the $x$-$y$ plane,
and similarly for $d_{xz}$ and $d_{yz}$. 
DFT~\cite{PRLSatpathy2008, JPSJImada2012} and ARPES experiments~\cite{NatureRozenberg2011} find
the in-plane hopping between $d_{xy}$ orbitals $t\!\simeq\!0.3$ eV, while
the out-of-plane $t'\!\simeq\!t/30$. The resulting $xz$ ($yz$) bands are quasi one-dimensional (1D),
dispersing primarily along $x$ ($y$), and confined along $z$.

{\bf Local moments:} To form moments, we need to localize charge, either through
disorder or correlations. Coupling to classical phonons is often argued to 
lead to localization, but quantum effects lead only to a bandwidth reduction;
localization needs disorder or correlations even in this case. 
Experiments \cite{AFetePRB2012} show that even the quasi-1D $xz, yz$ bands, which should 
be the most sensitive to disorder, continue to contribute to transport. This rules out strong 
Anderson localization.  

The key to understanding the effect of correlations is the splitting of the $t_{2g}$ degeneracy at the interface
described above. The 0.5e/Ti give rise to a quarter-filled $xy$ band in the top TiO$_2$ layer. 
We find that a modest on-site Hubbard  for $U$ and next-neighbor Coulomb $V$ then lead 
to a charge-ordered insulator (COI), see Fig.~\ref{Fig:1}(b).
We obtain a simple analytical result for the phase boundary between a metal and COI insulator
using a slave-rotor approach~\cite{PRBFlorens2004} (see Methods).
We also show that coupling to the breathing mode phonon further stabilizes the COI.
DFT calculations need a rather large $U \simeq 8$ eV to stabilize a COI~\cite{PRBPickett2008}. 
We see from Fig.~1 that, for realistic values~\cite{RMPImada} of $U = 4$ eV ($U/t \simeq 13$) and 
$V \simeq 0.5 - 1$ eV ($V/t \simeq 1.5 - 3$), we are deep 
in a checkerboard COI state: one sublattice occupied and the other empty.

\begin{figure}[!t]
\includegraphics[width=8cm]{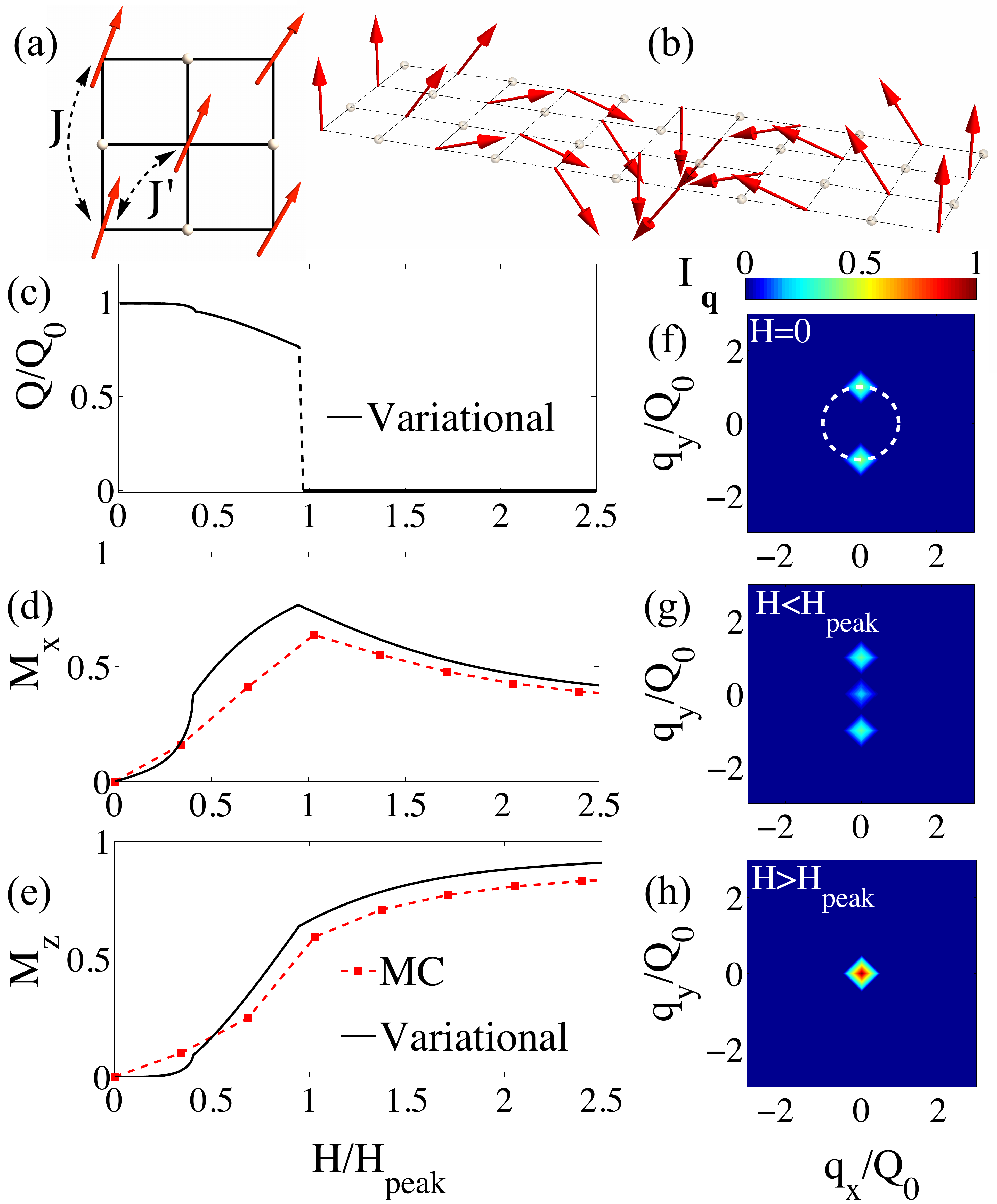}
\caption{{\bf Spiral ground state and its evolution with field:} 
(a) Local moments in the COI with FM exchange interactions $J$ and $J'$. The anisotropy and DM terms couple the same 
spins as $J$. 
(b) The zero-field spiral ground state, whose evolution with 
$\mb{H}=H(\sin\theta_H\hat{x}+\cos\theta_H\hat{z})$ with $\theta_H\!=\!15^\circ$
is shown next.
(c) The $H$-dependence of the spiral wavevector $Q$ normalized by its $H=0$ value $Q_0$
for SOC $\lambda/t=\pi/25$.
(d) In-plane ($M_x$) and (e) out-of-plane ($M_z$) components of the magnetization 
obtained  from $T\!=\!0$ variational calculations and low temperature ($T=0.25J$) Monte Carlo (MC) simulations. 
The $H$-axis is scaled by $H_\mr{peak}$ at which $M_x$ has a maximum. Panels
(f), (g), and (h) show the MC spin structure factor $I({\bf q})$. (f) At $H=0$ the peaks of the MC $I({\bf q})$
lie on a white circle, the locus of wavevectors $\mb{Q}_0$ corresponding to degenerate variational ground states.
(g) For $0< H<H_\mr{peak}$, the $I({\bf q})$ shows a net FM moment in addition to spiral peaks.
(h) A FM state is obtained for $H>H_\mr{peak}$. 
\label{Fig:2}}
\end{figure}

\begin{figure}[!t]
\includegraphics[width=8cm]{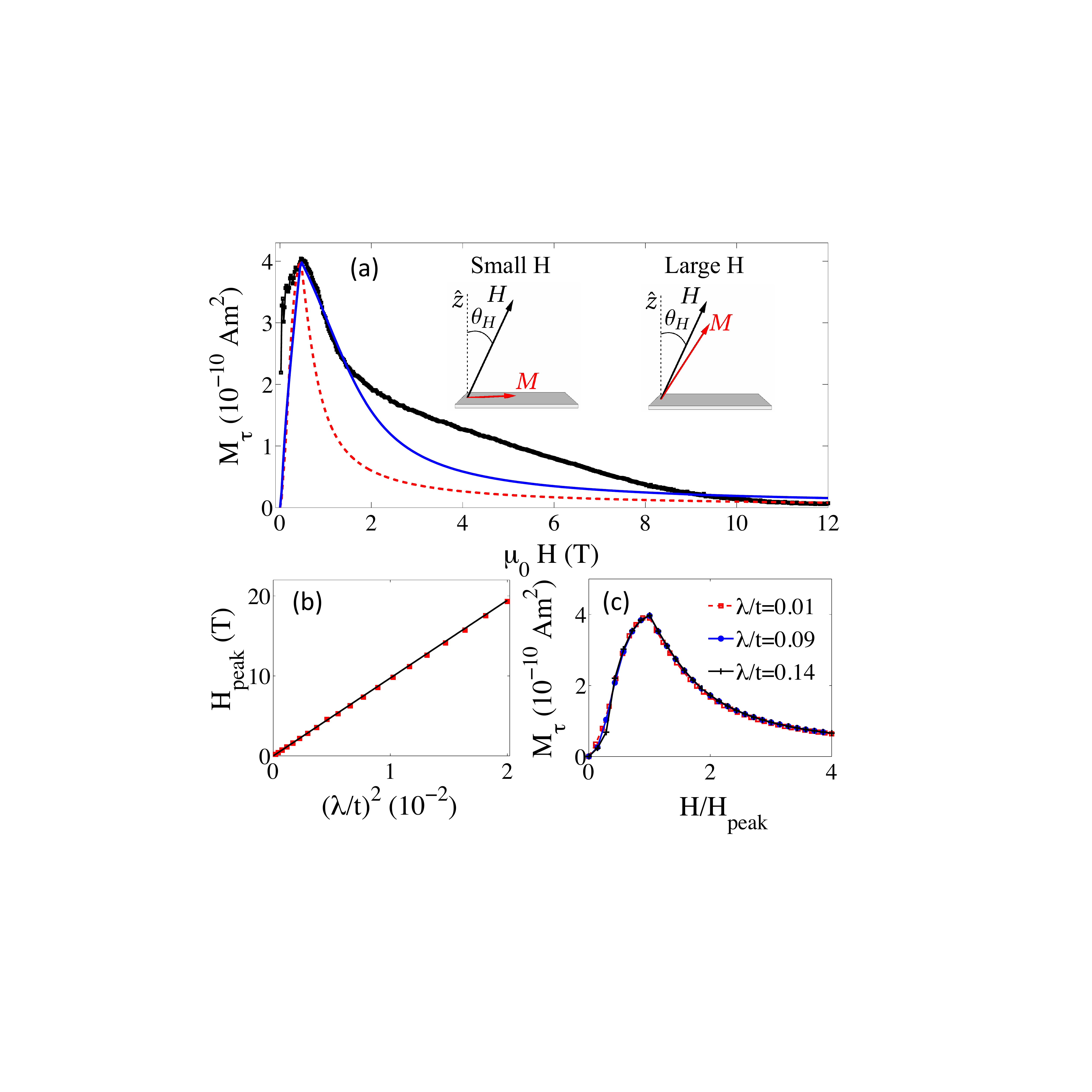}
\caption{{\bf Torque magnetometry:} (a) Magetization $M_\tau\!=\!\tau/H$ (black symbols) from torque experiments~\cite{NatPhyAshoori2011} 
at $T\!=\!300$ mK with $\theta_H\!=\!15^\circ$ shown in inset; (see Supplementary Information for background subtraction). 
Theoretical $M_\tau(H)$ curve (red dashed line) computed with microscopic value $AJ/D^2=1/2$
with $A\!\simeq\!0.3$ T chosen to match $H_\mr{peak}$.
Better agreement (blue line) with experiment is obtained by the phenomenological choice $AJ/D^2=0.8$ with $A\simeq 0.7$ T;
(see text). 
Inset shows schematic of two regimes in torque: an in-plane net magnetization $\mb{M}$ for small $H$ evolving into a
out-of-plane $\mb{M}$ for large $H$. (b) $H_\mr{peak}$ scales with $(\lambda/t)^2$ for fixed $n_c$, as predicted by the microscopic model. 
(c) $M_\tau(H)$ curves, for different values of SOC that can be tuned by gating, collapse onto a single curve when $H$ is scaled by $H_\mr{peak}$.
\label{Fig:3}}
\end{figure} 

{\bf Exchange:} We have now established a COI with $S=1/2$ local moments.
As argued in the introduction, symmetry then dictates the form of the effective magnetic Hamiltonian
$\mc{H}_\mr{eff}$. While we can make progress at a phenomenological
level, it is useful to ask what further insights a microscopic analysis leads to.

Consider then the problem of local moments in $d_{xy}$ orbitals on a checkerboard lattice [Fig.~\ref{Fig:2}(a)]
interacting via Hund's coupling $J_\mr{H}$ with a small density
$n_c$ of conduction electrons in $xz$, $yz$ bands (possibly coming from doping by oxygen vacancies).
Given $J_\mr{H}\simeq 1$ eV, it is much more reasonable to work in the non-perturbative double-exchange limit 
$J_\mr{H}\gg t$ than in the perturbative RKKY limit $J_\mr{H}\ll t$. 
We treat the $S=1/2$ moments classically. This is justified a posteriori because of the
small quantum fluctuations in the magnetic states -- FM or long-wavelength spiral --
that we analyze. 
 
We find $\mc{H}_\mr{eff}=\sum_\mb{r}\left[\mc{H}_x(\mb{r})+\mc{H}_y(\mb{r})\right]+\mc{H}'$ with
\begin{equation}
\mc{H}_x(\mb{r})=- J \mb{S}_\mb{r}\!\cdot\!\mb{S}_{\mb{r}+2\hat{x}} - A S_\mb{r}^y S_{\mb{r}+2\hat{x}}^y
- D\hat{y}\!\cdot\!(\mb{S}_\mb{r}\!\times\!\mb{S}_{\mb{r}+2\hat{x}})
\label{eq:effectiveH}
\end{equation}  
where $\mb{r}$ denotes positions of occupied sites in the COI; see Methods.  
$\mc{H}_y(\mb{r})$ is obtained by interchanging $x\leftrightarrow y$ and replacing $D\rightarrow -D$ in $\mc{H}_x(\mb{r})$.
The term $\mc{H}'=-J'\sum_\mb{r}\mb{S}_\mb{r}\cdot(\mb{S}_{\mb{r}+\hat{x}+\hat{y}}+\mb{S}_{\mb{r}+\hat{x}-\hat{y}})$
couples the nearest neighbour spins along the diagonal direction [Fig.~\ref{Fig:2}(a)]. 
The spins have normalization $\mb{S}_\mb{r}^2=1$.
The form of $\mc{H}_x$ and $\mc{H}_y$, as well as $J\gg J'$, arises from the 
two-sublattice structure of the COI and the quasi-1D nature of $xz$, $yz$ bands.
We find $J\simeq n_ct/4$ and $J'\simeq n_ct'/4$ for low carrier density $n_c$ (the number of electrons per Ti in each band).

The novel aspect of this double exchange model is the Rashba SOC 
$2\lambda a\hat{z}\!\cdot\!(\mb{k}\!\times\!{\boldsymbol \sigma})$ for the 
conduction electrons, where $\mb{k}$ is their momentum and 
${\boldsymbol \sigma}$ are Pauli matrices. 
The SOC strength $\lambda$ is determined by the electric field at the interface and 
transport experiments \cite{PRLCaviglia2010} find a
gate voltage tunable $\lambda\simeq 1.3$$-$$6.3$ meV, so that $\lambda/t\simeq 0.004$$-$$0.022$. 
SOC then gives rise to the DM and compass terms
in $\mc{H}_\mr{eff}$ via the double exchange mechanism.
For $\lambda/t\ll 1$, we find $D\simeq n_c\lambda$, $A \simeq 2n_c\lambda^2/t$ and $A'=0$. Two points are worth
noting. The ratio $AJ/D^2=1/2$ is independent of all microscopic parameters.
Remarkably, our double-exchange results $D/J\!\sim\!\lambda/t$ and $A/J\!\sim\!(\lambda/t)^2$ are
in the end very similar to Moriya's results~\cite{PRMoriya1960} for superexchange.

{\bf Spiral ground state:} We have examined the ground state properties of $\mc{H}_\mr{eff}$ in an external field
${\bf H}$ using a variety of analytical and numerical techniques. Let us first discuss $H\!=\!0$.
Using a variational approach we find that for $AJ/D^2\!<\!1$ the zero-field ground state is a spiral
\begin{equation}
{\bf S}_\mb{r} = \sin({\bf Q_0}\cdot{\bf r}){\bf \hat{Q}_0} + \cos({\bf Q_0}\cdot{\bf r})\hat{z}.
\label{spiral_solution}
\end{equation}
with ${\bf Q_0} = (2\lambda/ta)(\cos\varphi\,\hat{x} + \sin\varphi\,\hat{y})$ where $a$ is the Ti-Ti distance; see Methods.
For $AJ/D^2>1$, the ground state is an in-plane FM with $\mb{S}_\mb{r}=\cos\varphi\,\hat{x} + \sin\varphi\,\hat{y}$. 
For all $AJ/D^2$ , the ground state is infinitely degenerate (with arbitrary $0\!\leq\!\varphi\!<\! 2\pi$), 
even though $\mc{H}_\mr{eff}$ has $\mathbb{Z}_4$ symmetry.
This peculiar degeneracy of compass anisotropy~\cite{EPLBrink2004} is
expected to be broken by ``order-by-disorder''. 

The microscopically derived value $AJ/D^2=1/2$ implies a zero-field spiral variational ground state. We have verified 
the stability of the spiral over FM at $H\!=\!0$ using two independent, unbiased numerical calculations (see Supplementary Information): 
finite-temperature MC and $T\!=\!0$ conjugate gradient energy minimization.
In Fig.~\ref{Fig:2}(b) we show an example of spiral ordering in real space with a wavelength of a few lattice spacings.
For realistic values of $\lambda/t\sim 0.02$, our variational results give a pitch of $(2\pi/Q_0)\sim 600~\AA$.

In Figs.~\ref{Fig:2}(c)-(h), we show the evolution of the spiral state for $\lambda/t=\pi/25$ 
as a function of ${\bf H}$ applied in the $xz$-plane at an angle $\theta_\mr{H}=15^\circ$ with $z$-axis,
(the geometry used in the torque magnetometry experiments \cite{NatPhyAshoori2011}). 
In the variational calculation [Fig.~\ref{Fig:2}(c)] the spiral state (${\bf Q}\neq 0$) develops into a fully magnetized state above a field $H_\mr{peak}$, 
which corresponds to a maximum in the in-plane component $M_x$ of the magnetization 
$\mb{M}=\langle \mb{S}_\mb{r}\rangle$. We plot in Figs.~\ref{Fig:2}(d,e) $M_x$ and $M_z$ as functions of $H$,
where we also see excellent agreement between the variational and the unbiased MC results.
Here we used larger SOC that led to smaller pitch spirals for the MC simulations.
Below we use the variational approach to study long wavelength spirals for realistic SOC $\lambda$.

The evolution from the spiral state to the FM as a function of field is best seen in the MC
spin structure factor $I_\mb{q}\propto |\langle \mb{S}_\mb{q}\rangle|^2$ shown in Figs.~\ref{Fig:2}(f)-(h),
where $\mb{S}_\mb{q}$ is the Fourier transform of $\mb{S}_\mb{r}$. The coplanar spiral at $H=0$ shows two peaks in
$I_\mb{q}$ at $\mb{q}=\pm Q_0\hat{y}$ [Fig.~\ref{Fig:2}(f)], which lie on the
white circle, the locus of $\mb{Q}_0$ corresponding to the degenerate ground states
in eq.~(\ref{spiral_solution}). At intermediate fields, a uniform component develops along with the non-zero $\mb{q}$ peaks [Fig.~\ref{Fig:2}(g)],
while for $H>H_\mr{peak}$ one obtains a FM state [Fig.~\ref{Fig:2}(h)]. 

{\bf Torque and magnetization:} 
The torque ${\boldsymbol \tau}=\mb{M}\times\mb{H}$ is non-zero only in the presence of 
spin-space anisotropy, otherwise $\mb{M}$ would simply align with $\mb{H}$. 
In Fig.~\ref{Fig:3}(a), we compare our variational results with the magnetization $M_\tau \equiv \tau/H$ {\it vs.}~$H$ derived 
from torque magnetometry~\cite{NatPhyAshoori2011}, where $M_\tau$ is the component of $\mb{M}$ perpendicular to $\mb{H}$.
(The experimental data have been shown after background subtraction; see Supplementary Information.)

We can understand $M_\tau(H)$ as follows. In our theory, an external magnetic field with an 
in-plane component ($\theta_H\!\neq\!0$) first induces a uniform magnetization in the spiral state [Fig.~\ref{Fig:2}]. 
This leads to an increasing $M_\tau$ for $H\!<\!H_\mr{peak}$ in Fig.~\ref{Fig:3}(a), where $H_\mr{peak}$, at which $M_\tau$ is a maximum, 
depends on $\theta_H$. For $H\!>\!H_\mr{peak}$, the ground state is a FM [Fig.~\ref{Fig:2}], in which the DM term is irrelevant, 
but the compass term gives rise to a large easy-plane anistropy $A\simeq 0.15-0.3$ T (with $\lambda\simeq 0.016-0.022$ and $n_c=0.05$) 
that tries to keep $\mb{M}$ in the plane. For large enough $H$ the field dominates over the anisotropy, pulls $\mb{M}$ out of the plane 
and $M_\tau$ decreases. These two regimes are shown schematically in Fig.~\ref{Fig:3}(a) inset. 
  
In Fig.~\ref{Fig:3}(a), we show two theoretical $M(H)$ curves which differ in their choice of $AJ/D^2$. 
The red curve corresponds to the Rashba value $AJ/D^2=1/2$, with $A= 2n_c\lambda^2/t=\!0.3\!$ tuned to match $H_\mr{peak}$
(with $\lambda/t=0.022$ and $n_c\!=\!0.05$).
The high field drop in $M_\tau(H)$ is much too rapid compared with the data. In the blue curve, we choose
$AJ/D^2=0.8$ with $\lambda/t=0.028$ which leads to better agreement with experiment. 
Such a phenomenological choice of $AJ/D^2$ amounts to changing the planar anisotropy 
while keeping the DM interaction fixed. In our microscopic theory, both $A$ and $D$ were determined by Rashba SOC. 
In addition, several other effects also contribute to $A$, including dipolar interactions and atomic SOC (see Supplementary Information).

The peak position $H_\mr{peak}\!\propto\!n_c\lambda^2/t$ in our microscopic model and $M_\tau(H)$ 
for different $\lambda$ collapse onto a single curve when plotted {\it versus} $H/H_\mr{peak}$; see Figs.~\ref{Fig:3}(b),(c). 
In practice, other sources of anisotropy might modify this scaling. Nevertheless, both $n_c$ and $\lambda$ are gate tunable 
and we expect $H_\mr{peak}$ to vary substantially with bias.

{\bf Ferromagnetic patches:} We now discuss how we can reconcile the torque data~\cite{NatPhyAshoori2011}
discussed above with the scanning SQUID results of ref.~\cite{NatPhyMoler2011}.
A detailed modeling of inhomogeneity seen in ref.~\cite{NatPhyMoler2011} is beyond the scope of this paper, nevertheless, we can offer a 
plausible picture based on our theory.  At $H\!=\!0$ most of the sample has, in fact, coplanar spiral order and 
hence no net magnetization. However, the energy gain of the spiral over FM order, 
$\Delta\epsilon\simeq (D^2/J-A)/2\!=\!n_c\lambda^2/t\!\simeq\!0.1\!-\!0.2$~K, is quite small; see Methods. 
Therefore small terms ignored up to now might well upset the balance in favor of FM. In particular,
micron size patches could be possible if the coupling to strain fields was involved.
  
We thus look at the effect of magnetoelastic coupling associated with polar distortion, a novel aspect of these interfaces. 
The Rashba SOC $\lambda(u)$ is determined by local electric fields, which are affected by  
small displacements $u\hat{z}$ of oxygen ions that bridge the Ti-O-Ti bonds~\cite{arXivMacDonald2013}.  Both $A$ and $D$ terms in 
$\mc{H}_\mr{eff}$ try to maximize energy gain by increasing $\lambda(u)$ by a displacement $u<0$
with $\lambda(u)\!=\!\lambda\!-\!\lambda_1u$ where $\lambda_1>0$; see Methods. But this costs an electrostatic 
energy $-qEu$, where $\mb{E}\!=\!-E\hat{z}$ is the electric field acting on the charge $q$ of the oxygen ion, in addition to an elastic cost
$Ku^2/2$ with elastic constant $K$. Minimizing the total energy with respect to $u$, we find that the 
balance between spiral and FM can be reversed; see Methods.
We hypothesize that strain influences the microscopic parameters in such a way that it stabilizes FM patches,
which can then point in any direction in the plane, consistent with compass anisotropy.

{\bf Discussion:} We comment briefly on how our work, which builds on insights 
from electronic structure calculations \cite{PRBPickett2006, PRBPickett2008, PRLSatpathy2008, JPSJImada2012}, differs from other theoretical approaches. 
It is hard to see how one can get the large exchange coupling or net moment seen in LAO/STO from
itinerant models~\cite{arXivBalents2013,NJPFischer2013}, which may well be relevant for other interfaces like GdTiO$_3$/SrTiO$_3$.
Our model differs from ref.~\cite{PRLLee2012} in several respects. 
We are in the non-perturbative double-exchange limit, as noted above, and not the RKKY regime
that leads to much smaller exchange. More importantly, in our model, $xz,yz$ carriers mediate exchange.
The $xy$ electrons in lower layers have essentially no interaction with moments in the $xy$ orbitals of the top TiO$_2$ layer.
This results from the small overlap $t'$ between ${xy}$ orbitals along $\hat{z}$ as well as the level mismatch due to confinement. 
Conventional superconductivity arises in $xy$ states in lower layers and is decoupled from the local moments. 

{\bf Conclusions:}
Our theory has several testable predictions. Local moment formation is associated with $(\pi,\pi)$ charge order at the interface. The  
zero-field state is predicted to be a coplanar spiral state with several striking properties, including a wavevector that scales linearly with
Rashba SOC and hence is gate tunable. The evolution from a spiral to a FM state in an external field has characteristic signatures
in the spin structure factor. The easy-plane anisotropy strength is also dominated by SOC and exhibits substantial gate voltage dependence
that can be tested in torque magnetometry experiments. The FM exchange $J = n_ct/4\simeq 40$ K (for $n_c = 0.05$ and 
$t= 0.3$ eV) should also be tunable by changing $n_c$, the density of carriers.
A theoretical analysis of the finite temperature properties of our model is left for future investigations.

{\bf Acknowledgments}: We gratefully acknowledge stimulating conversations with L. Li, W. Cole, J. Mannhart, K. Moler,
W. Pickett, S. Satpathy and N. Trivedi, and the support of NSF-DMR-1006532 (O.E., M.R.) and DOE-BES DE-SC0005035 (S.B.).

\bigskip\bigskip\bigskip
\noindent
\underline{\bf Methods}

{\bf Effective Spin Hamiltonian:} Broken inversion along $\hat{z}$
and in-plane square-lattice symmetry constrains the
terms arising from SOC. In $\sum_{{\bf r},\mu}\hat{d}_\mu\!\cdot\!({\bf S}_{\bf r}\!\times\!{\bf S}_{{\bf r}+b\hat{\mu}})$ 
the only allowed DM vector is $\hat{d}_\mu = \hat{z}\!\times\!\hat{\mu}$. Other choices like
$\hat{d}_\mu\!=\!\hat{z}$ break $\pi/2$ rotation and $\hat{d}_x\!=\!\pm\hat{x}, \hat{d}_y\!=\!\pm\hat{y}$ break
in-plane reflection. Similarly, only the compass terms shown in the text are permitted. Terms that involve
$S^x S^y$ or  $S^xS^z$ or $S^yS^z$ are not allowed by symmetry.

{\bf Charge ordering Mott transition:} We consider the quarter-filled, extended Hubbard model on a 2D square lattice
with onsite $U$ and nearest-neighbor Coulomb $V$.  
Using slave-rotor~\cite{PRBFlorens2004} mean field theory (MFT) for $U$ and Hartree-Fock for $V$,
we obtain an analytical result for the transition from metal to charge ordered insulator (COI).
The checkerboard COI is stable for
$U>8\langle t \rangle$ when $U<4V$, and for $(U-2V)V/U > \langle t \rangle$ when $U>4V$.
Here $\langle t \rangle \simeq 0.66t$ is the kinetic energy of occupied states.
Slave rotor MFT has been found to be in excellent semi-quantitative agreement with
dynamical mean field theory (DMFT) for several problems.
In fact, our 2D square lattice results are quite similar to DMFT of the 
Wigner-Mott transition on a Bethe-lattice~\cite{NatPhysKotliar2008}. 

The COI is stabilized by coupling to the breathing mode phonon,
which has the same $(\pi,\pi)$ wavevector as the charge order; 
see Supplementary Information. The only change above is that 
$V \rightarrow \widetilde{V} = V+E_{ph}$, where $E_{ph}$ is the energy gained though lattice distortion.
Typical values of $E_{ph} \simeq 0.05-0.10 $eV \cite{PRBSatpathy2011}. 

{\bf Double Exchange:}
The local moments ${\bf S}_\mb{r}$ on the sites of a 2D checkerboard lattice are described by their
orientation $(\theta_\mb{r},\varphi_\mb{r})$. In the large $J_H$ limit, a conduction electron
on the ``local-moment sublattice'' ($a$) has its spin parallel to ${\bf S}_\mb{r}$. Thus we write the 
spin-full fermion operators $\tilde{a}_{\mb{r}\alpha\sigma}$, with orbital index $\alpha\!=\!(xz),(yz)$, 
in terms of spinless fermions $a_{\mb{r}\alpha}$,
via $\tilde{a}_{\mb{r}\alpha\sigma}^\dagger\!\rightarrow\!F_{r\sigma} a_{r\alpha}^\dagger$.
Here $F_{\mb{r}\uparrow}\!=\!\cos(\theta_\mb{r}/2)$ and $F_{\mb{r}\downarrow}\!=\!\sin(\theta_\mb{r}/2)e^{-i\varphi_\mb{r}}$. 

Both spin projections $\sigma\!=\!\uparrow,\downarrow$ are allowed on ``empty sublattice'' ($b$) sites,
for which we use a common (global) quantization axis. The kinetic energy is then given by
\begin{equation}
{\cal H}_K = -\sum_{\mb{r},\alpha, \hat{\mu}, \sigma} t_{\alpha \hat{\mu}} F_{\mb{r}\sigma} a^\dagger_{\mb{r}\alpha}(b_{\mb{r}+\hat{\mu},\alpha \sigma} + b_{\mb{r}-\hat{\mu},\alpha \sigma})+\mr{h.c.}
\label{eq:kineticH}
\end{equation}
where $t_{(xz),\hat{x}}\!=\!t_{(yz),\hat{y}}\!=\!t$ and $t_{(xz),\hat{y}}\!=\!t_{(yz),\hat{x}}\!=\!t'\simeq t/30$.

The Rashba terms $\sigma^x \sin(k_y a )\!\sim\!-k_ya({\boldsymbol \sigma} \times \hat{z})_y$ 
and $\sigma^y \sin(k_x a )\!\sim\!k_xa({\boldsymbol \sigma} \times \hat{z})_x$ lead to the SOC Hamiltonian
\begin{eqnarray}
{\cal H}_{R}&=&i\lambda \sum_{\mb{r},ij} \left[\sigma^x_{ij} \tilde{a}^\dagger_{\mb{r}(yz)i}(b_{\mb{r}+\hat{y},(yz)j}-b_{\mb{r}-\hat{y},(yz)j})\right.\nonumber \\
&&\left.-\sigma^y_{ij} \tilde{a}^\dagger_{\mb{r}(xz)i}(b_{\mb{r}+\hat{x},(xz)j}-b_{\mb{r}-\hat{x},(xz)j}) \right] + {\rm h.c.} 
 \end{eqnarray}
We can rewrite this in terms of the spinless $a$'s as 
\begin{equation}
{\cal H}_{R} = \sum_{\mb{r}\alpha \sigma} \Big[ \lambda_{\alpha\sigma} F_{\mb{r} \overline{\sigma}} a^\dagger_{\mb{r}\alpha}(b_{\mb{r}+\hat{\mu},\alpha \sigma}- 
b_{\mb{r}-\hat{\mu},\alpha \sigma})\Big] + {\rm h.c.} 
\label{eq:rashbaH}
\end{equation}
where $\overline{\sigma}\!=\!-\sigma$ and we define $\lambda_{(xz)\uparrow}\!=\!-\lambda_{(xz)\downarrow}\!=\!-\lambda$ and 
$\lambda_{(yz)\uparrow}\!=\!\lambda_{(yz)\downarrow}\!=\!i\lambda$.

To obtain the parameters $J$, $J'$, $D$ and $A$ of $\mc{H}_\mr{eff}$ of eq.~(\ref{eq:effectiveH})
starting from the microscopic $\mc{H}_\mr{DE}=\mc{H}_\mr{K}+\mc{H}_\mr{R}$, we match the the energies of several 
low-lying configurations computed for both $\mc{H}_\mr{DE}$ and $\mc{H}_\mr{eff}$.
See Supplementary Information for details.

{\bf Variational calculation:} We study the ground state properties of $\mc{H}_\mr{eff}$ using a variational ansatz 
$\mb{S}_\mb{r}=\mb{M}+\mb{R}(\mb{Q})\cos(\mb{Q}\cdot\mb{r})
+\mb{I}(\mb{Q})\sin(\mb{Q}\cdot \mb{r})$ with $\mb{Q} \neq 0$. 
The normalization ${\bf S}_r^2\!=\!1$ is satisfied via the constraints 
${\bf M}\cdot{\bf R} = {\bf M}\cdot{\bf I} = {\bf R}\cdot{\bf I}= 0$; ${\bf R}^2 = {\bf I}^2$ and ${\bf M}^2 +{\bf R}^2 = 1$. 
We numerically minimize the energy per spin $\epsilon$  to obtain optimal values 
for $\mb{M}$, $\mb{R}$, $\mb{I}$ and $\mb{Q}$ as a function of $\mb{H}$,
and calculate the torque.  

At zero field, we find a spiral ground state of eq.~(\ref{spiral_solution}) with $\mb{Q}\!=\!\mb{Q}_0$  and $\mb{M}\!=\!0$ for $AJ/D^2\!<\!1$ 
and an in-plane FM for $AJ/D^2>1$. 
We can see this analytically using the ansatz $\mb{S}_\mb{r}=\mb{R}\cos(\mb{Q}\cdot\mb{r})
+\mb{I}\sin(\mb{Q}\cdot\mb{r})$.
We write $\mb{R}=\sin\theta\sin\varphi\hat{x}-\sin\theta\cos\varphi\hat{y}+\cos\theta\hat{z}$ and
$\mb{I}=\cos\varphi\hat{x}+\sin\varphi\hat{y}$ with $\theta\in[0,\pi]$ and $\varphi\in[0,2\pi)$. 
We first minimize with respect to $\mb{Q}$ (for $Qa\ll 1$), find
the optimal $\mb{Q}\!=\!(D/2Ja)\cos\theta(\cos\varphi\hat{x}+\sin\varphi\hat{y})$
and the energy $\epsilon(\mb{Q})\simeq -2(J+J')-A+(A-{D^2}/{J})(\cos^2{\theta})/2$, 
which is independent of $\varphi$. For $AJ/D^2<1$, the optimal choice of $\cos\theta=1$ leads to 
the spiral of eq.~(\ref{spiral_solution}) with $\mb{Q}\!=\!\mb{Q}_0$ 
and an energy gain $\Delta\epsilon=(D^2/J-A)/2$ relative to the FM state. For our microscopic 
model $\Delta\epsilon=n_c\lambda^2/t$.

The effect of a polar distortion $u\hat{z}$ of oxygen ions on the SOC can be modeled by 
replacing $A\!\rightarrow\!A(u)=2n_c\lambda^2(u)/t$ and $D\!\rightarrow\!D(u)=n_c\lambda(u)$
in $\epsilon(\mb{Q})$ above, in addition to the energy cost $(Ku^2/2-qEu)$. 
Here $K$ is the elastic constant for out of plane distortion at the interface and
$\mb{E}=-E\hat{z}$ is the electric field acting on oxygen ions at the interface.
$\lambda(u)\!=\!\lambda\!-\!\lambda_1u$ with $\lambda_1>0$ since the
local electric field, and hence SOC, decreases if negatively charged oxygen ion moves up ($u\!>\!0$).
We minimize the total energy with respect to $u$ to find that 
energy gain of the spiral is reduced:
 $\Delta\epsilon\simeq (n_c\lambda^2/t)[1-2qE\lambda_1/K\lambda]$. 
With suitable choice of parameters, we can make  $\Delta\epsilon\!<\!0$ and stabilize the FM state over the spiral.

\newpage
\section{Supplementary Information}

{\bf 1) Electron-phonon coupling in charge-ordered insulating state:}
Here we show how the charge-ordered insulator (COI) is further stabilized by coupling to the breathing mode phonon. The electron-phonon coupling can be written as $H_{ep}=(K_0/2)\sum_i (Q_{ai}^2+Q_{bi}^2)-g\sum_i (Q_{ai}n_{ai}+Q_{bi}n_{bi})$. 
Here $a$, $b$ label the two sublattices of the COI in the $i$-th unit cell, $K_0$ is the elastic constant,
$g$ the electron-phonon coupling strength for breathing model displacements $Q$'s of Ti ions, and $n_{ai}$'s are the electronic occupancies. 
In the COI state with one sublattice occupied $n_{ai}=1$ and the other empty $n_{bi}=0$, we take the staggered displacement pattern 
$Q_{ai}=-Q_{bi}=Q$ and minimize the energy with respect to $Q$. The energy gained though such lattice distortion is $E_{ph} =g^2/4K_0$, 
which has a typical value of $\simeq 0.05-0.10 $eV \cite{SPRBSatpathy2011}. Incorporating the effects of $H_{ep}$ into
our slave rotor analysis, we find that the results are the same as those described in the Methods with the replacement $V \rightarrow \widetilde{V} = V+E_{ph}$. 
This enhancement in $V$ is due to the cooperative distortion of the breathing mode as the charge order have the same $(\pi,\pi)$ wavevector.

\bigskip

{\bf 2) Microscopic derivation of parameters for $\mc{H}_\mr{eff}$:}
We obtain the parameters $J$, $J'$, $D$ and $A$ of $\mc{H}_\mr{eff}$ 
by comparing the energies of several low-lying configurations with that obtained 
from the double exchange Hamiltonian 
$\mc{H}_\mr{DE}=\mc{H}_\mr{K}+\mc{H}_\mr{R}$ described in the Methods. Here we briefly sketch the procedure.

Consider an $N\times N$ (with even $N$)  lattice of Ti sites with 
$N^2/2$ local moments in a checkerboard arrangement (see Fig.~2(a) in the main text).
Consider conduction electrons in the low-density limit $n_c\ll 1$ so that only the states near the bottom 
of the $xz$ and $yz$ bands are filled.

We first neglect $t'\ll t$ and take strictly 1D $xz$ and $yz$ bands. 
This leads to decoupled 1D channels along $x$ and $y$. For $t'=0$, there is no coupling ($J'=0$) between 
nearest-neighbor spins of the COI along diagonal direction. The system then essentially splits into two inter-penetrating, 
decoupled subsystems. (We will consider the $t'\neq0$ coupling between these subsystems below.)
Let us consider the energies of the following two configurations of local moments: (a) uniform FM state and (b) a twisted spin configuration.
 
(a) An uniform FM with $\theta_\mb{r}=\theta$ and $\varphi_\mb{r}=\varphi$. 
We diagonalize $\mc{H}_\mr{DE}$ to obtain the lowest-lying energy bands for $xz$ and $yz$ carriers, given by
\begin{equation*}
e_{xz}(\mb{k})\simeq -2t+t(k_x-\frac{\lambda}{t}\sin\theta\sin\phi)^2-\frac{\lambda^2}{t}\sin^2\theta\sin^2\varphi,
\end{equation*}
and $e_{yz}(\mb{k})$ obtained by replacing $k_x\rightarrow k_y$ and $\sin\varphi\rightarrow \cos\varphi$. Depending on the spin configuration the system gains energy by shifting the band bottoms. This essentially leads to the easy-plane anisotropy as can be seen by computing the total energy per spin to leading order in $n_c$ and $\lambda/t$, namely, $\epsilon_\mr{DE}(\theta,\varphi)\simeq -8n_ct-(2n_c\lambda^2/t)\sin^2\theta$. This can be compared with 
the energy of the same configuration in the spin-model, $\epsilon_\mr{eff}(\theta,\varphi)\simeq -2J-A\sin^2\theta$ to obtain the 
result $A=2n_c\lambda^2/t$ mentioned in the main text. 
(The angle-independent constant determines the absolute value of energy and is not a meaningful quantity to 
compare between microscopic and low-energy effective models.) 

 (b) A twisted spin configuration along $x$ direction with $\theta_{\mb{r}+2\hat{x}}-\theta_\mb{r}=\delta\theta\rightarrow 0$, $\theta_{\mb{r}+2\hat{y}}=\theta_\mb{r}$ and $\varphi_\mb{r}=0$. 
In this case the $xz$ band gives rise to $N$ identical 1D channels at $n_y=1,..,N$. One can analytically diagonalize the $xz$ part of the Hamiltonian, by transforming the $b$-site fermions into new rotated fermions $f$'s as 
\begin{subequations} \label{eq.b_rotation}
\begin{eqnarray}
b_{\mb{r}+\hat{x},xz\uparrow}&=&\cos\left(\theta_\mb{r}/2\right)f_{\mb{r}+\hat{x},xz\uparrow}
-\sin\left(\theta_\mb{r}/2\right)
f_{\mb{r}+\hat{x},xz\downarrow},\\
b_{\mb{r}+\hat{x},xz\downarrow}&=&e^{i\varphi_\mb{r}}[\sin\left(\theta_\mb{r}/2\right)f_{\mb{r}+\hat{x},xz\uparrow}
+\cos\left(\theta_\mb{r}/2\right)
f_{\mb{r}+\hat{x},xz\downarrow}],\nonumber
\\&&
\end{eqnarray}
\end{subequations} 
provided $(N/2)\delta\theta<\pi$ so that $0\leq\theta_\mb{r}<\pi$ and $\varphi_\mb{r}=0$ $\forall \mb{r}$.
The energy per spin from this part turns out to be $\epsilon_{xz}(\delta\theta)\simeq -4n_ct-n_c\lambda\delta\theta+(n_ct/8)\delta\theta^2$. The $yz$ band leads to $N$ decoupled 1D channels at $n_x=1,..,N$, but each with a separate FM configuration of spins with angles $\theta=n_x\delta\theta/2$ and $\varphi=0$. Here we can use the result for $e_{yz}(\mb{k})$ obtained in (a) and calculate the energy contribution per spin from $yz$ band 
as $\epsilon_{yz}(\delta\theta)\simeq -4n_ct-2n_c\lambda^2/t$. The total energy for this configuration is $\epsilon_\mr{DE}(\delta\theta)\simeq -8n_ct-2n_c\lambda^2/t-n_c\lambda\delta\theta+(n_ct/8)\delta\theta^2$ compared with that obtained from $\mc{H}_\mr{eff}$, $\epsilon_\mr{eff}(\delta\theta)\simeq -2J-A-D\delta\theta+(J/2)\delta\theta^2$. This gives $J=n_ct/4$ and $D=n_c\lambda$.  

Finally,  we switch on $t'\neq 0$ and obtain $J'$ (see Fig.~2(a) in the main text). 
Here we neglect the small SOC terms (again of the DM and compass form) that are suppressed by a factor $(t'/t)$ relative to $D$ and $A$. 
We now consider a twisted spin configuration (c) 
where $\theta_{\mb{r}+\hat{x}+\hat{y}}-\theta_\mb{r}=\delta\theta$, $\theta_{\mb{r}+2\hat{x}}=\theta_{\mb{r}+2\hat{y}}=\theta_\mb{r}$ and $\varphi_\mb{r}=\varphi$. The fermion Hamiltonian can be diagonalized by rotating $b$-site fermions via equation \eqref{eq.b_rotation}, now for both $xz$ and $yz$ carriers. The energy in this case to leading order in $t'/t$ is $\epsilon_\mr{DE}(\delta\theta)\simeq -8n_c(t+t')+(n_ct'/4)\delta\theta^2$. The corresponding energy from spin model is $\epsilon_\mr{eff}(\delta\theta)\simeq -2J'+J'\delta\theta^2$. This leads to a $J'=n_ct'/4\ll J$.

{\bf 3) Monte Carlo and numerical minimization:} 
We have performed Monte Carlo (MC) simulation \cite{SNewmanMC} and conjugate gradient minimization \cite{SPressNR} as unbiased checks 
on the results obtained from variational calculation for ground state properties of $\mc{H}_\mr{eff}$. 
In order to access spiral wavevector $Q_0=(2\lambda/ta)$, we have taken commensurate values of $\lambda=\pi/p$ with $p=25,30,50$ 
for $2\times np\times np$ spins ($n=1,2$) on the lattice of Fig.~2(a) (main text). We use standard Metropolis sampling \cite{SNewmanMC} for 
MC simulation with unit length spins $\mb{S}_\mb{r}$ and perform $10^5$ MC steps per spin for equilibration, followed by $4\times 10^5$ MC steps/spin to calculate magnetization $\mb{M}=\langle \mb{S}_\mb{r}\rangle$ and spin structure factor $I_\mb{q}\propto |\langle \mb{S}_\mb{q}\rangle|^2$. The finite-temperature spin structure factor plotted in Figs.~2(f)-(h) is normalized with $I=\sum_\mr{BZ} I_\mb{q}$ summed over the first Brillouin zone.     

\bigskip

{\bf 4) Analysis of torque magnetometry data:} 
In the main text [Fig.~3(a)] we compare our theoretical results with the background subtracted torque magnetometry data \cite{SNatPhyAshoori2011}.
Here, we describe the background subtraction procedure used for the data and its rationale. We
also discuss the scaling of the magnitude of the magnetization.

Apart from $\mb{M}\times \mb{H}$ from the sample, the experimental torque ${\boldsymbol \tau}$ contains additional ``background'' contributions
that are apparent at high fields $H \apgt 5$ T. First there is a paramagnetic piece of the torque
$\tau_\mr{qu} \sim H^2$ that originates from the STO substrate \cite{SNatPhyAshoori2011}.
The raw data for the torque $\tau$ and the measured paramagnetic signal $\tau_\mr{qu}$ from the STO 
substrate (extrapolated up to high fields) from ref.~\cite{SNatPhyAshoori2011} are plotted in Supplementary Fig.~\ref{Fig:1S}(a).

We see from Supplementary Fig.~\ref{Fig:1S}(b) that the subtracted torque $(\tau-\tau_{qu})$ is a linear function  ($\tau_\mr{lin}$) of
$H$ at fields in excess of $10$ T, which in fact has a zero (or negligible) intercept when extrapolated down to
$H=0$. Why is the high field behavior linear in $H$ even beyond 15T? We consider two possibilities.
One possible explanation could be an extremely high, and rather unprecedented, strength of easy plane anisotropy $>15$ T.
In this case, the magnetization $\mb{M}$ vector is not pulled out of the plane of the sample to completely align with 
the external field even at 15 T, for only then could one get a torque linear in $H$. This seems to us most unlikely;
certainly none of the mechanisms we are aware of would account for such a large anisotropy.

An alternative explanation (Lu Li, private communication) is that the high field linear-$H$ signal arises
from a magnetic field inhomogeneity that gives rise to force $\left(\mb{M}\cdot{\nabla}\right) \mb{H}$. 
Given the fact that the high-field part is found to vary from sample to sample~\cite{SNatPhyAshoori2011}, 
we think that this second explanation is much more viable. 

We have therefore
subtracted this high-field linear background from the torque. In Fig.~3(a) (main text) we plot 
$M_\tau(H)=(\tau-\tau_{qu}-\tau_{lin})/H$ for the data of ref.~\cite{SNatPhyAshoori2011}.

In comparing our theory with the experimental $M_\tau(H)$ we have simply scaled our
$M_\tau(H_\mr{peak})$ to match the experimental value in Fig.~3(a) (main text). However, we point out that there is quite
reasonable agreement between theory and experiment even if he had not scaled our theory and plotted
the results in absolute units using a $g$-factor of 2. We do not try and derive the exact scaling factor microscopically in view of the
uncertainties in the actual $g$-factors of the local moments (0.5 per Ti) and of the conduction electrons
($2n_c$ per Ti from the two bands). Note that in the double-exchange model the conduction electron spins are
locked to those of the local moments.

\begin{figure}[!t]
\includegraphics[width=8cm]{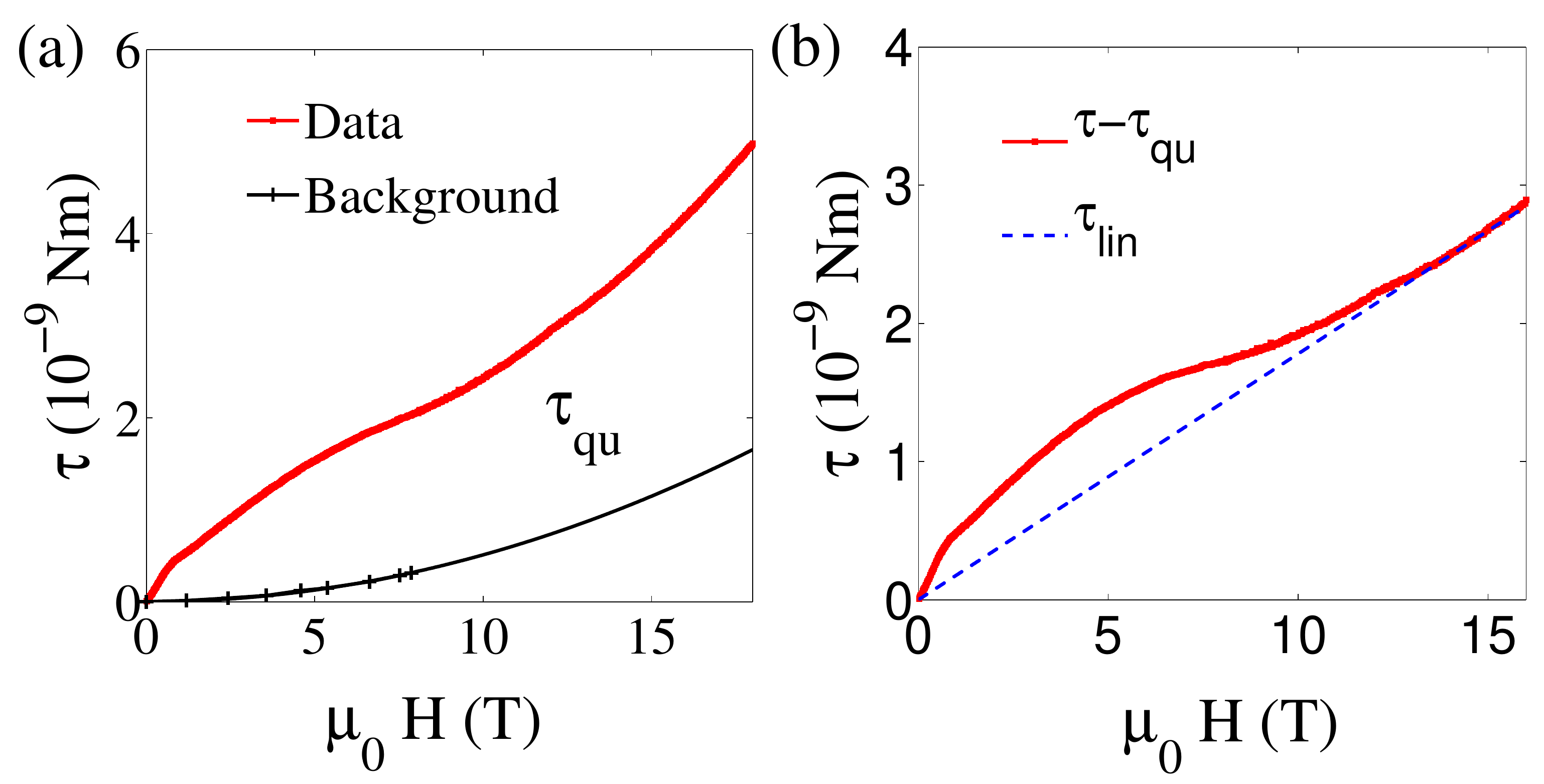}
\caption{{\bf Analysis of torque data:}(a) Torque magnetometry data for LAO/STO interface from ref.~\cite{SNatPhyAshoori2011}. Also shown is the quadratic background ($\tau_{qu}$) measured for STO substrate. (b) Shows the high field linear-$H$ contribution $\tau_\mr{li}$ extrapolating through the origin.} 
\label{Fig:1S}
\end{figure}

{\bf 5) Magnetic Anisotropy:} 
In the text we focussed on anisotropy arising from Rashba SOC.
Other mechanisms include shape anisotropy (via dipolar interactions) and atomic SOC~\cite{SRepProgJohnson1996}. 
Although these are likely to be subdominant to the Rashba contribution, as argued below,
nevertheless, the phenomenological fit in Fig.~3(a) (main text) suggests the importance of more easy-plane anisotropy than can be accounted for by Rashba SOC alone.

FM domains of size $\sim 1 \mu$m or larger would prefer to align in-plane in 2D or quasi-2D systems to minimize the 
long-range dipolar interaction. The energy contribution per unit volume due to such shape effect is \cite{SRepProgJohnson1996, SPRBDubowik1996} 
$\epsilon_\mr{dip}\simeq (\mu_0/2)\mb{m}^2\cos^2\theta$, where $\mb{m}$ is the average magnetization per unit volume and $\theta$ the angle it subtends with the normal to the plane of the interface. Phenomenologically, we can consider it as a term in the spin model of the form 
$A_s\sum_\mb{r}(S_\mb{r}^z)^2$, where the easy-plane anisotropy due to shape effect is 
$A_s= \mu_0 \mu_\mr{B}^2 S^2/4a^3$, with $S=1/2$ and $2a^3$ as the volume occupied by the individual local moments in the COI. 
As a result, $A_s \simeq 0.015$ T. Note that this crude estimate of shape anisotropy is much smaller than the compass anisotropy term 
$A\simeq 0.3$ T that arises from Rashba SOC. 

Atomic SOC prefers the magnetization to point along crystalline axes, leading to four-fold in-plane anisotropy. 
This must be small since the FM patches, seen in scanning SQUID experiments~ \cite{SNatPhyMoler2011},
point in random directions in the plane.

\end{document}